\documentclass[%
 reprint,
 amsmath,amssymb,
 aps,
 pre,
]{revtex4-2}

\usepackage{graphicx}
\usepackage{dcolumn}
\usepackage{bm}
\usepackage{hyperref}
\usepackage{multirow}
\usepackage{array}
\usepackage{ae} 
\usepackage{aecompl}
\usepackage{xcolor}
\usepackage{breqn} 

\begin{document}


\title{Emergent cell migration from cell shape deformations and T1 transitions}

\author{Harish P. Jain}
\affiliation{
Njord Centre\\ Department of Physics\\ University of Oslo}
 
\author{Richard D.J.G. Ho}
\affiliation{
Njord Centre\\ Department of Physics\\ University of Oslo}

\author{Luiza Angheluta}
\email{luiza.angheluta@fys.uio.no}
\affiliation{
Njord Centre\\ Department of Physics\\ University of Oslo}

\begin{abstract}
    T1 transitions, which are localised cell rearrangements, play an important role in fluidization of epithelial monolayers. Using a multiphase field model and an active elastic solid model, we show that although each cell undergoes T1 transitions in time as uncorrelated, random events, the spatial distribution of these events is highly correlated and is dependent on cell shape. T1 transitions have a dual effect. Cells losing neighbours tend to relax their shape, while those gaining neighbours tend to elongate. By analysing the statistics of successive T1 transitions undergone by a deformable cell, we find asymmetric spatial distributions related to how cells lose or gain neighbours. These asymmetric spatial patterns of T1 transitions promote directed cell migration, and form the backbone for coherent flow patterns at tissue scales.
\end{abstract}
\maketitle

\section{Introduction}
Confluent epithelial monolayers, as composed of tightly packed cells with stable cell-cell junctions, exhibit a complex interplay between tissue rigidity, collective migration, and fluidization~\cite{Ladoux2017, hakimCollectiveCellMigration2017}. These mechanical processes are also important in assisting biological phenomena such as morphogenesis, cancer invasion, and wound healing~\cite{Friedl2009, bruguesForcesDrivingEpithelial2014}. Cells within a tissue monolayer exhibit collective migration either by moving together as coherent clusters or through neighbour rearrangements~\cite{Ladoux2017}. Even when cells move as coherent clusters, the cells at the outer edges of these clusters undergo neighbour exchanges, such as T1 transitions, with nearby cells. While cells within the monolayer may also undergo T1 transitions when they move at different velocities. 

T1 transitions, observed in active cell monolayers~\cite{irvineCellIntercalationDrosophila1994} and in foams under external forces~\cite{weaireSoapCellsStatistics1984a}, are fundamental topological transitions. During one T1 transition, a junction between two neighbouring cells shrinks and disappears, while two adjacent cells form a new junction typically oriented perpendicular to the original one. T1 transitions result in changes in neighbourhood connectivity without any change in cell number. Although the geometric features of T1 transitions are well understood, the underlying mechanisms triggering these events vary significantly. In passive systems such as soap films, T1 transitions are driven by external forces~\cite{stavansEvolutionCellularStructures1993}. However, for biological tissues, these transitions are influenced by intrinsic cellular properties, such as active cell migration~\cite{jainRobustStatisticalProperties2023, jainCellIntercalationFlow2024a}, or mechano-chemical feedback involving actomyosin and junctional protein complexes such as a adherens junctions~\cite{bertetMyosindependentJunctionRemodelling2004, blankenshipMulticellularRosetteFormation2006}.

Despite their importance, many aspects of T1 transitions remain poorly understood. Although these events are thought to dissipate junctional stress and mediate solid-fluid transitions in tissues~\cite{mongeraFluidtosolidJammingTransition2018, kimEmbryonicTissuesActive2021a, mongeraMechanicsCellularMicroenvironment2023a}, the underlying processes are not yet fully resolved.

T1 transitions sustained by cell activity induce large deformations in cell shape, associated with an energy barrier~\cite{bi2014energy, jainRobustStatisticalProperties2023}. Recently, we have shown that T1 transitions triggered by cell activity induce transient quadrupolar flows, leading to the relative dispersion of cells~\cite{jainCellIntercalationFlow2024a}. Interestingly, the rate of relative dispersion is determined by the mean occurrence rate of T1 transitions, regardless of the underlying mechanisms controlling how T1 transitions may occur. However, the spatiotemporal patterns of T1 transitions, hypothesized as chains or cascades of T1 transitions~\cite{rosaNucleationGlideDislocations1998, jainRobustStatisticalProperties2023} propagating across the tissue to induce tissue fluidization and coherent flow structures on larger scales, remain elusive.

In this paper, we study how T1 transitions driven by cell activity contribute to large deformations in cell shape, which in turn influence where next T1 transitions are likely to occur. To explore the general properties of T1 transitions as neighbour exchange events and as junction remodelling events that dependent on cell shapes, we emply two complementary modeling approaches: the multi-phase field (MPF) model and the active elastic solid (AES) model. 

In the MPF model, cells are represented as deformable entities with shape boundaries tracked by a phase field. MPF models have been used to study several emergent properties of epithelial monolayers, such as topological features including T1 transitions~\cite{wenzelTopologicalGeometricalQuantities2019, jainRobustStatisticalProperties2023, jainCellIntercalationFlow2024a}, p-atic order~\cite{mueller2019emergence, Saw2017, armengol-colladoEpitheliaAreMultiscale2023}, and cell colony growth~\cite{nonomuraStudyMulticellularSystems2012, jainImpactContactInhibition2022}. 

Conversely, the AES model represents cells as point-like entities that are self-propelled and connected by elastic springs~\cite{ferrante2013elasticity}. This model has been used to reproduce observations of polar ordering in epithelial monolayers~\cite{laang2024topology}. We enhance this model to enable neighbour exchanges based on a geometric condition ~\cite{ho2024role}. 

In both modeling approaches, we find that each cell undergoes T1 transitions as a Poisson process over time, with the occurrence rate dependent on the cell's self-propulsion activity. This suggests that the Poissonian nature of T1 transitions is likely a property of topological changes in the cell neighbour graph. Additionally, the properties of cell junctions significantly influence the spatial distribution of T1 transitions, as these depend on the feedback between cell shape deformations and T1 transitions. The MPF model is particularly useful for exploring how this feedback leads to a highly asymmetric distribution of successive T1 transitions across cell junctions.

In Section~\ref{sec:T1_transitions} we describe T1 transitions in MPF and AES models. Within these models, we demonstrate that T1 transitions, as temporal events triggered by cell activity, follow a Poisson process with a mean rate determined by activity. This is discussed in Section~\ref{Sec:waiting_time}. By measuring the radial distribution functions, we find that T1 transitions exhibit spatial correlations that decay over time, consistent with the hypothesis of chaining of T1 transitions, as discussed in Section~\ref{sec:spatial distribution of T1 events}. In Section~\ref{sec: Shape changes}, we show that T1 transitions induce distinct shape deformations depending on whether a cell gains or loses neighbours. Cells losing neighbours undergo shape relaxation, while those gaining neighbours experience elongation. In Section~\ref{sec: successive T1s}, through the analysis of successive T1 transitions undergone by a cell, we demonstrate that the spatial distribution of T1 transitions is non-uniform and depends on specific topological changes. This spatial non-uniformity leads to the emergence of directional migration. Finally, in Section~\ref{Sec:conclusions}, we summarize the results and provide an outlook on the topic.

\section{T1 transitions}\label{sec:T1_transitions}
\begin{figure*}[]
    \centering 
    \includegraphics[width=\textwidth]{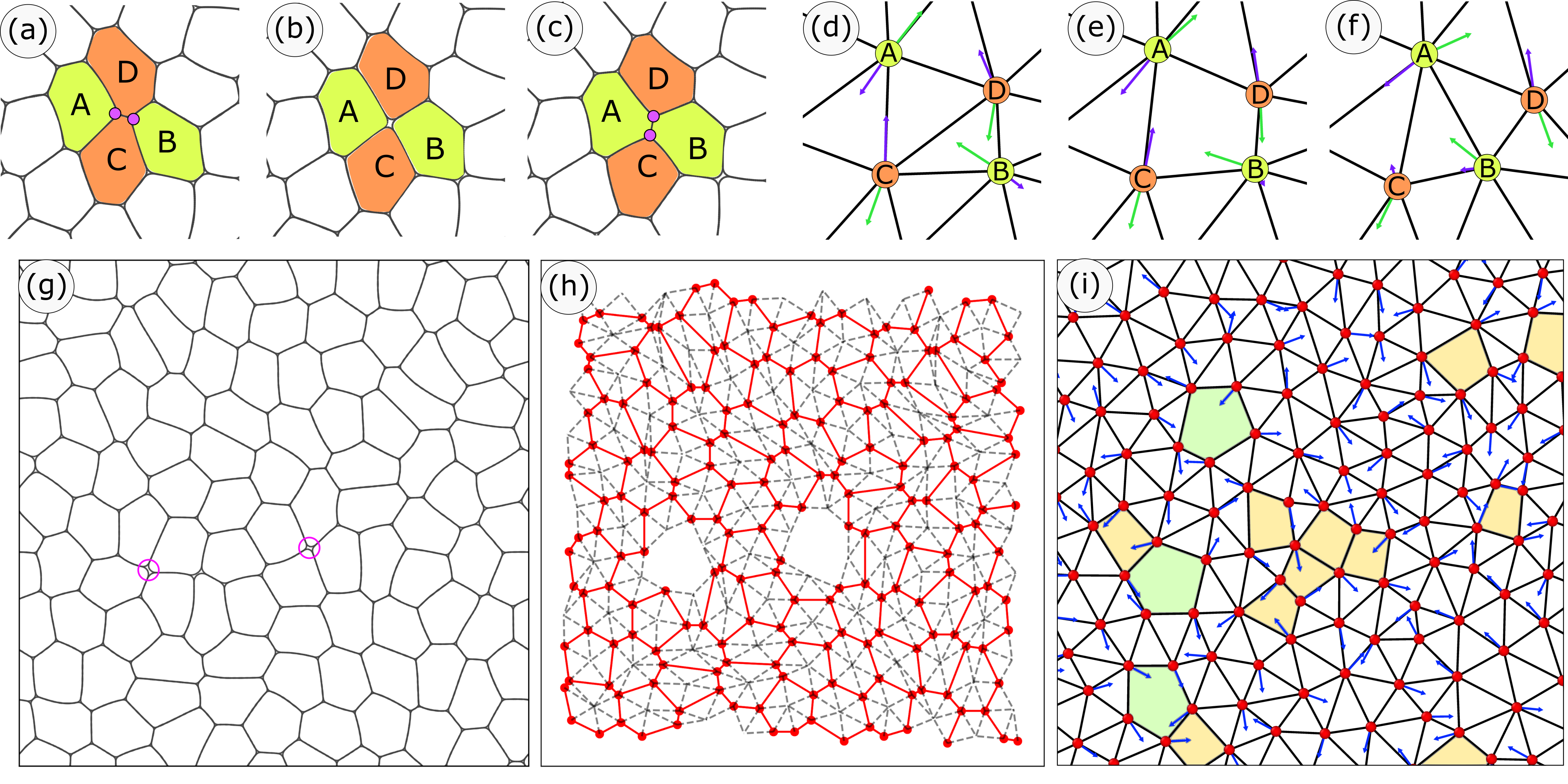}
    \caption{T1 transition in the MPF model (a-c) and in the AES model (d-f). (a) and (d) before T1; (b) and (e) during T1; (c) and (f) after T1. A T1 transition $\mathcal{T}=\{A,B,C,D\}$ where cells $A$ and $B$ are loser cells, while cells $C$ and $D$ are gainer cells. (a-c) pink markers corresponds to location of vertices that take part in the T1 transition. (d-f) AES: Black line correspond to cell-cell junctions, orange and green circles mark the loser and gainer cells, respectively. 
    The green arrows represents the cell polarity, i.e. direction of self-propulsion, whilst the purple arrow represents the net elastic force. (g) A snapshot of a tissue in the MPF model with the pink circle marking the location of the T1 transitions. (h) The vertices is marked in red, and vertices that are common to two cells are linked by red lines. The black dash lines connect the vertices to the center of mass of the cells. The gaps in the plot corresponds to T1 transitions. (i) Tissue snapshot of AES model. Cells are marked in red, and black bar connects neighbouring cells. Triangles indicate vertices, yellow trapeziums indicates T1 transitions and green polygons denote higher order rosettes. See Movie 1 and Movie 2 in the  Supplemental Material for simulations in MPF model and AES model, respectively. }
    \label{fig:1}
\end{figure*}

We consider a 2D confluent monolayer of $N$ cells, with their index labels in the set $\mathcal{M} = \{1, 2, \dots, N\}$.  The neighbour connectivity is expressed in terms of a symmetric neighbour adjacency matrix $\mathcal{N}_{N \times N}$ at time $t$ as
\begin{equation*}
    \mathcal{N}_{ij}(t) = \left\{ 
     \begin{array}{rl}
     1 &: \;\; i,j \in \mathcal{M} \text{ are neighbours at t and } i\neq j \\ 0 &: \;\; \text{otherwise} 
     \end{array}\right.
\end{equation*}
In a multi-phase field (MPF) model, each cell $i$ is represented by a phase field $\phi_i$ defined such that $\phi_i \approx 1$ inside the cell and $\phi_i \approx -1$ outside it. The cell boundary is characterized by a thin diffuse region where the phase field transitions smoothly between these two bulk values. Such a representation allows us to model cells of arbitrary shapes. The evolution of the phase field is governed by diffusive relaxation aimed at minimizing a free energy functional, with an advection term due to cell self-propulsion. The mathematical formulation of the model is described in Appendix~\ref{App:MPF} and has been previously used in Reference~\cite{jainCellIntercalationFlow2024a}. MPF model parameters are as per Table~\ref{tab:parameters} unless specified otherwise. The conservative dynamics preserves the total cell area. 

A T1 transition, denoted by $\mathcal{T}$, is a neighbour-exchange event of finite duration that involves four cells $A, B, C, D$. $\mathcal{T}$ starts at time $t=t_{\mathcal{T}}^{-}$ when cells $C$ and $D$ lose contact and concludes at $t=t_{\mathcal{T}}^{+}$ when cells $A$ and $B$ establish contact. This rearrangement process is illustrated in Figure~\ref{fig:1} (panel a-c). Before $t_{\mathcal{T}}^{-}$, the junction shared by cells C and D shrinks and vanishes at $t_{\mathcal{T}}^{-}$, and a new junction between A and B, approximately perpendicular to the previous one, is created at $t_{\mathcal{T}}^{+}$ and subsequently starts to lengthen~\cite{jainCellIntercalationFlow2024a}. Thus, T1 transitions in MPF models result in the remodelling of cell-cell junctions.  This T1 transition can be represented by the set $\mathcal{T} = \{A, B, C, D\} \subset \mathcal{M}$  such that 
\begin{eqnarray*}
  \mathcal{N}_{CD}(t_{\mathcal{T}}^{-}) = 1 \Rightarrow \mathcal{N}_{CD}(t_{\mathcal{T}}^{+}) = 0\\
  \mathcal{N}_{AB}(t_{\mathcal{T}}^{-}) = 0 \Rightarrow \mathcal{N}_{AB}(t_{\mathcal{T}}^{+}) = 1,
\end{eqnarray*}
while all other connections shared among the four cells remain unchanged during the T1 duration $\Delta t_{\mathcal{T}}= t_{\mathcal{T}}^+ - t_{\mathcal{T}}^-$. Cells $A$ and $B$, which gain neighbours during the T1 transition, are referred to as \emph{gainer} cells, while cells $C$ and $D$, which lose neighbours, are referred to as \emph{loser} cells. During the T1 transition, an extracellular gap is formed between the four cells in the MPF model, as shown in the Figure~\ref{fig:1} (panel b). While neighbour exchanges involving more than four cells, forming higher-order rosette structures, are possible in the MPF model, they are extremely rare events in the explored parameter regime and thus their influence is ignored. 

A 3-way vertex denoted by the subset $\mathcal{X}=\{i, j, k\} \subset \mathcal{M}$, is a set of three cells which are mutually adjacent, i.e.,  $\mathcal{N}_{ij}(t) = \mathcal{N}_{jk}(t) = \mathcal{N}_{ki}(t) = 1$, at a given time $t$. The set of all vertices at time $t$ is denoted by $\Sigma_{\mathcal{X}}(t)$. For a unique $\mathcal{N}(t)$, there exists a unique $\Sigma_{\mathcal{X}}(t)$, which represents a particular neighbourhood topology. T1 transitions are topological changes of $\mathcal{N}$ (or equivalently $\Sigma_{\mathcal{X}}$). From hereon, we will refer to a 3-way vertex as just a vertex. 

We determine the position of the vertex $\mathcal{X}$ at time $t$, denoted by $\mathbf{r}_{\mathcal{X}}(t)$, as the location of the minimum value of the sum of its distances to boundaries of all three cells. The boundary of a cell $i$ is implicitly represented by the contour corresponding to $\phi=0$, and is represented by the zero-level set $Z_i(t)$ such that
\begin{equation}
    Z_{i}(t)=\{\mathbf{x} \in \Omega \;|\; \phi_i(\mathbf{x}, t) = 0\}.
\end{equation} The position of the vertex $\mathcal{X}$ is then given as
\begin{equation}
    \mathbf{r}_{\mathcal{X}}(t) = \arg\min_{\mathbf{x}} \left\{ \sum_{j \in \mathcal{X}}|\mathbf{x}-\mathbf{z}_j| : \forall \mathbf{z}_j \in Z_j(t) \right\}
\end{equation}
In Figure~\hyperref[fig:1]{1g}, we show a typical snapshot of a cell configuration in MPF model, while in the Figure~\hyperref[fig:1]{1h}, we plot the graph of the corresponding vertices. For each T1 transition $\mathcal{T}$, at time $t_{\mathcal{T}}^{-}$, two vertices $\mathcal{X}_{\mathcal{T}, 1}^- =\{A, C, D\}$ and $\mathcal{X}_{\mathcal{T}, 2}^- =\{B, C, D\}\subset \mathcal{T}$ move toward each other and annihilate (Figure~\hyperref[fig:1]{1a}). Similarly, at $t_{\mathcal{T}}^{+}$,  two new vertices $\mathcal{X}_{\mathcal{T}, 1}^+ =\{A, B, C\}$ and $ \mathcal{X}_{\mathcal{T}, 2}^+ =\{A, B, D\}\subset \mathcal{T}$ are spawned (Figure~\hyperref[fig:1]{1c}). Notice that the small gaps formed between cells during T1 transitions correspond to holes in the \emph{graph of vertices} as shown in Figure~\hyperref[fig:1]{1h}. 

As localised processes of remodelling cell-cell junctions, T1 transitions induce large fluctuations in cell shape and speed, generating transient flows with quadrupolar patterns that result in an increase in the pair-separation distance between cells~\cite{jainRobustStatisticalProperties2023, jainCellIntercalationFlow2024a}. 

To better understand the interplay between cell shapes and T1 transitions, we also consider an active elastic solid (AES) model where the tissue is idealized as an elastic sheet formed by pointwise cells which are self-propelled and interact with each other through elastic forces. The deterministic evolution of the polarity for the self-propulsion is determined by the torque induced by misalignments between the net elastic force and the cell polarity. Although this model features tissue elasticity, it is not directly related to interfacial forces as is the case for the MPF model. The model is detailed in Appendix~\ref{App:ASM} and based on the one introduced in Reference~\cite{ferrante2013elasticity}. AES model parameters are as per Table~\ref{tab:parameters} unless specified otherwise.

In the AES model, the neighbouring cells are initially connected by a predefined cell connectivity graph. In previous works, this graph was quenched in time, resulting in a solid-like response when the tissue is out of its elastic equilibria, or a polar collective migration due to alignment of polarities~\cite{ferrante2013elasticity}. We augment this model by introducing a geometric relation of nearest neighbours which is the same as the one proposed in Reference~\cite{nissen2018theoretical,ho2024role} to allow for neighbour exchanges. 
Explicitly, we connect all cells $i$ and $j$ and remove the connection if there exists any third cell $k$ which satisfies $r_{ij}^2 > r_{ik}^2 + r_{jk}^2$, with $r_{ij}$ being the distance between cells $i$ and $j$.
Most of the neighbour exhanges correspond to T1 configurations as shown in Figure~\ref{fig:1} (panels d,e,f) and marked by the yellow quadrilateral shapes in the connectivity graph (see panel i). However, neighbour exchanges involving five cells are also present, as illustrated by pentagonal shapes in panel (i). 

For the AES model, note that the T1 transitions are topological changes in the \emph{connectivity graph} and are not directly related to 3-way vertices as in the MPF model. In the MPF model, two cells are considered neighbours if their shape boundaries sufficiently overlap, such that topology changes in the connectivity graphs corresponds in junction remodeling. In the AES model, cells are considered neighbours if they conform to a specific geometric configuration of cell positions. This geometric dependence of neighbourhood topology allows for neighbour exchanges with four or more cells, with neighbour exchanges involving four cells being the most common. By contrast, in MPF models, neighbour exchanges almost always involve four cells within the parameter regime explored. For simplicity, we also use the symbol $\mathcal{T}$ to denote neighbour exchanges in the AES model, even if they involve more than four cells. In the AES model, for our analysis, neighbour exchanges involving more than four cells are counted as one transition, and the algorithm for detecting neighbour exchanges is described in Appendix~\ref{App:ASM}.

In the AES model, when the self-propulsion is high, the cells are driven far out of their equilibrium positions imposed by elastic interactions, resulting highly non-uniform distribution of distances between particles, analogous to that of a gas, which is unrealistic limit case for epithelia.
Therefore, the AES with neighbour exchanges is valid at sufficiently small activities, where there is still a well-defined reference state for elastic deformations, which is only punctually destroyed by neighbour exchanges. Thus, we restrict ourselves to situations with moderate self-propulsion speeds. Thus, both models have a connectivity graph, encoded within $\mathcal{N}$, where T1 transitions occur as localised neighbour exchange events.

\section{Waiting-time between T1 Transitions}\label{Sec:waiting_time}

\begin{figure}
    \centering
    \includegraphics[width=1.0\linewidth]{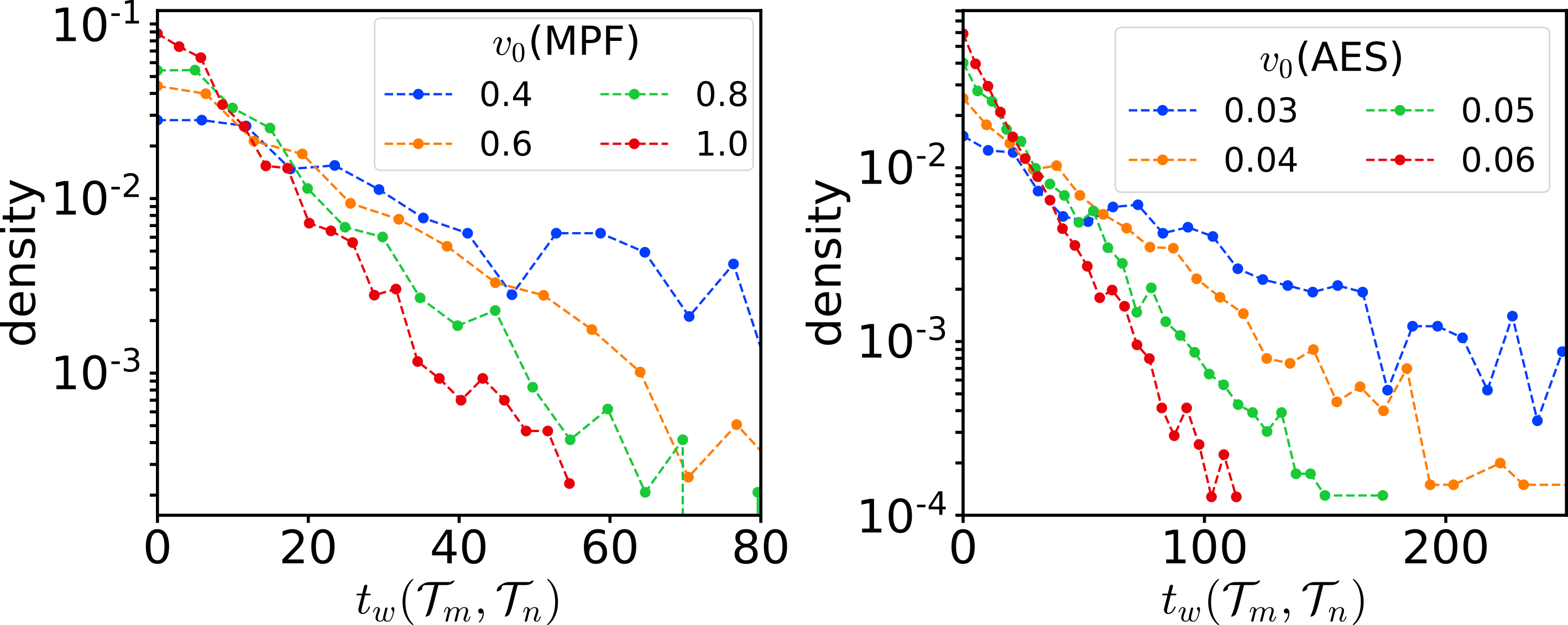}
    \caption{Waiting time $t_w$ probability density distribution between \emph{successive} T1s $\mathcal{T}_m$ and $\mathcal{T}_n$ undergone by a cell. (a) MPF model: cell activity $v_0$ is varied. (b) AES model: cell activity $v_0$ is varied. }
    \label{fig:2}
\end{figure}

To quantify the spatiotemporal relations between T1 transitions, we introduce the notion of two \emph{successive} T1 transitions $\mathcal{T}_m$ and $\mathcal{T}_n$ involving cell $i$ i.e. $i\in \mathcal{T}_m$ and $i\in \mathcal{T}_n$. We consider two T1s involving a cell $i$, \emph{successive} if $t_{\mathcal{T}_m}^{+} < t_{\mathcal{T}_n}^{-}$, and if cell $i$ is not part of any other T1 between $t_{\mathcal{T}_m}^{+}$ and $t_{\mathcal{T}_n}^{-}$. 
The waiting time $t_w(\mathcal{T}_m, \mathcal{T}_n)$ between \emph{successive} T1 transitions $\mathcal{T}_m$ and $\mathcal{T}_n$ is defined as the time interval between end of $\mathcal{T}_m$ and start of $\mathcal{T}_n$ i.e., $t_w(\mathcal{T}_m, \mathcal{T}_n) := t_{\mathcal{T}_n}^{-} - t_{\mathcal{T}_m}^{+}$. 

Figure~\hyperref[fig:2]{2a} and Figure~\hyperref[fig:2]{2b} show the distributions of waiting times $t_w$  for varying activity ($v_0$) in the MPF and AES models, respectively. In both cases, the waiting time is exponentially distributed. The average waiting time decreases with increasing activity as cells that move faster are likely to undergo their next T1 sooner. The mean waiting time does not vary significantly with varying deformability $Ca$ (see Figure~1 in Supplemental Material). Cells with lower deformability typically undergo fewer T1 transitions; however, these T1 transitions have a larger duration~\cite{jainRobustStatisticalProperties2023}. 

Whilst T1 transitions appear Poissonian in time, we show next that T1 transitions are correlated in space. Later, we will use the cell shape of the MPF to describe the spatial correlations in the MPF model.

\section{Spatial correlation of T1 Transitions}\label{sec:spatial distribution of T1 events}

\begin{figure*}
    \centering
    \includegraphics[width=\textwidth]{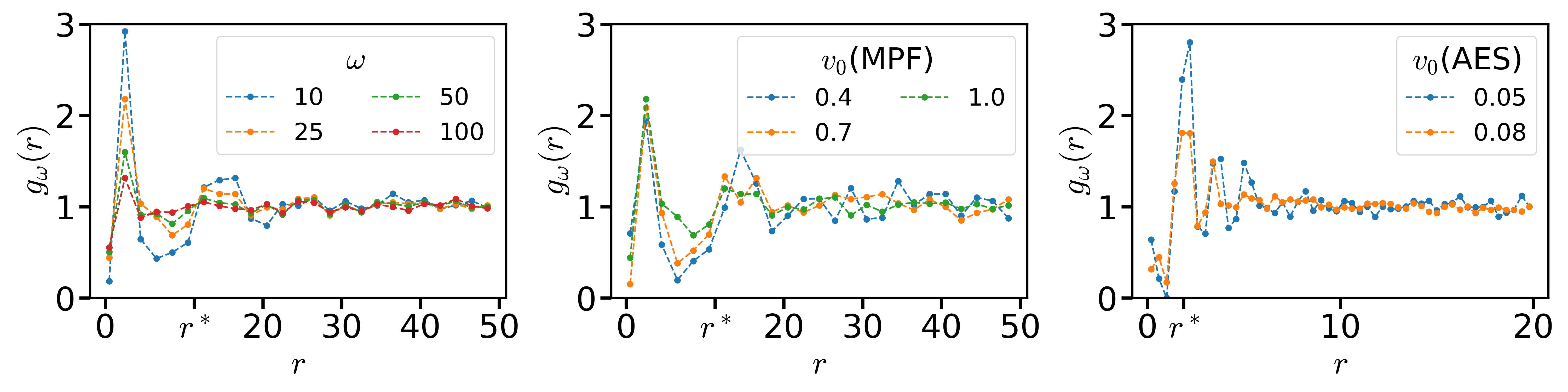}
    \caption{Radial distribution function $g_{\omega}(r)$ between T1 transitions for MPF model (a-b) and AES model(c). (a) $g_{\omega}(r)$ for different time windows for $v_0=1$. (b) $g_{\omega}(r)$ for varying $v_0$ with $w=25$. (c) $g_{\omega}(r)$ for varying $v_0$ with $w=10$. $r^*$ is the mean separation distance.}
    \label{fig:3}
\end{figure*}

To study the spatial correlations between T1 events, we compute the radial distribution function for pairs of T1 events. We first compute the epicenters of T1 transitions. Within MPF model, the epicenter $\mathbf{r}_{\mathcal{T}}$ of a T1 transition $\mathcal{T}$ is empirically defined as the location where the sum of distances to the boundaries of all four cells involved in the T1 transition is minimized. This is measured at time midway through the T1 transition, i.e., at $t_{\mathcal{T}}^m= 0.5(t_{\mathcal{T}}^+ + t_{\mathcal{T}}^-)$~\cite{jainRobustStatisticalProperties2023, jainCellIntercalationFlow2024a}. The location of epicenter is given by
\begin{equation}
    \mathbf{r}_{\mathcal{T}}(t) = \arg\min_{\mathbf{x}} \left\{ \sum_{j \in \mathcal{T}}|\mathbf{x}-\mathbf{z}_j| : \forall \mathbf{z}_j \in Z_j(t_{\mathcal{T}}^m) \right\}
\end{equation}
The epicenter lies within the extracellular gap created during the T1 transition, as shown in Figure~\hyperref[fig:1]{1b}. In the AES model, the epicenter is defined as the average of the positions of the four cells involved in $\mathcal{T}$, i.e.
\begin{equation}    
\mathbf{r}_{\mathcal{T}}(t_m) = \frac{\mathbf{r}_{A} +  \mathbf{r}_{B} +  \mathbf{r}_{C} +  \mathbf{r}_{D}}{4}
\end{equation}
with time taken at the end of the T1 transition, $t_{\mathcal{T}}^m= t_{\mathcal{T}}^+$. 

For a specific time window of size $\omega$, we compute the radial distribution function for epicenters of all T1 transitions that occur within that window.  Suppose, in a given time window of size $\omega$ starting from time $t$ until time $t+\omega$, there are $N_{T1}$ events and let $\rho=\frac{N_{T1}}{A}$ be the number density of T1 events, where $A$ is the area of the computational domain. Then, the radial distribution function for this particular time window is calculated as 
\begin{equation}
    g(r, t, t+\omega) = \frac{1}{\rho N_{T1}}\left< \sum_{j=1}^{N_{T1}} \sum_{k \neq j} \delta(r - r_{jk}) \right>
\end{equation}
where $\delta$ is the Dirac delta function and $r_{jk}$ is the distance between the epicenters of a pair of T1 transitions indexed by $j$ and $k$. If $\tau$ is the time resolution, then the radial distribution function is calculated as the ensemble average of time windows, $\{(0, \omega), (\tau, \omega+\tau), (2\tau, \omega+2\tau), \dots, , (T-\omega, T)\}$, such that 
\begin{equation}
    g_{\omega}(r) = \frac{\tau}{(T-\omega+\tau)}\sum_{t=0}^{t=T-\omega}g(r, t, t+\omega)
\end{equation}
Figure~\hyperref[fig:3]{3a} shows the radial distribution function for the MPF model for varying time window sizes $\omega$. The location of the peaks does not vary with window size, but decrease in magnitude. For small $\omega$, $g_{\omega}(t)$ appears typical of disordered systems with interacting particles. As epicenters of T1 transitions lie in the vicinity of cell-cell junctions, the peaks in the $g_{\omega}(t)$ must be relate to where the cell-cell junctions are located. Over time, the junctions remodel due to T1 transitions, and so $g_{\omega}(t)$ flattens as $\omega$ increases. Figure~\hyperref[fig:3]{3b} shows $g_{25}(t)$ for varying activity $v_0$ in MPF model. As activity increases, the correlations reduce due to higher rates of junction remodelling~\cite{jainCellIntercalationFlow2024a}. Similar results are observed for the AES model, as shown in Figure~\hyperref[fig:3]{3a}, where $g_{\omega}(t)$ is plotted for varying activity $v_0$. 

The peaks in the AES model lie at units of the mean separation distance $r^*$, determined by the number density of cells. In the MPF model, the cell separation distance $r^*$ is defined as the diameter of the circle whose area is equal to the cell's area. By contrast to the AES model, the first peak lies at $r\approx 3$, followed by a valley at $r\approx 8$ and then another peak at $r\approx 15$, all within the mean cell separation $r^*$. Thus, the location of the first two peaks and the first valley in $g_{\omega}(t)$ for low $\omega$ correspond to different cell junctions across a cell. We discussed this further in Section~\ref{sec: successive T1s} by analyzing spatial distributions of \emph{successive} T1 transitions undertaken by a cell.

From here on, we make use of the high shape resolution of cells in the MPF model to discuss the interplay between cell shape and T1 transitions. The shape will also help us better understand the profile of the radial distribution $g_{\omega}(r)$ of MPF model.

\section{Shape changes induced by individual T1 events}\label{sec: Shape changes}

\begin{figure*}
    \centering
    \includegraphics[width=\textwidth]{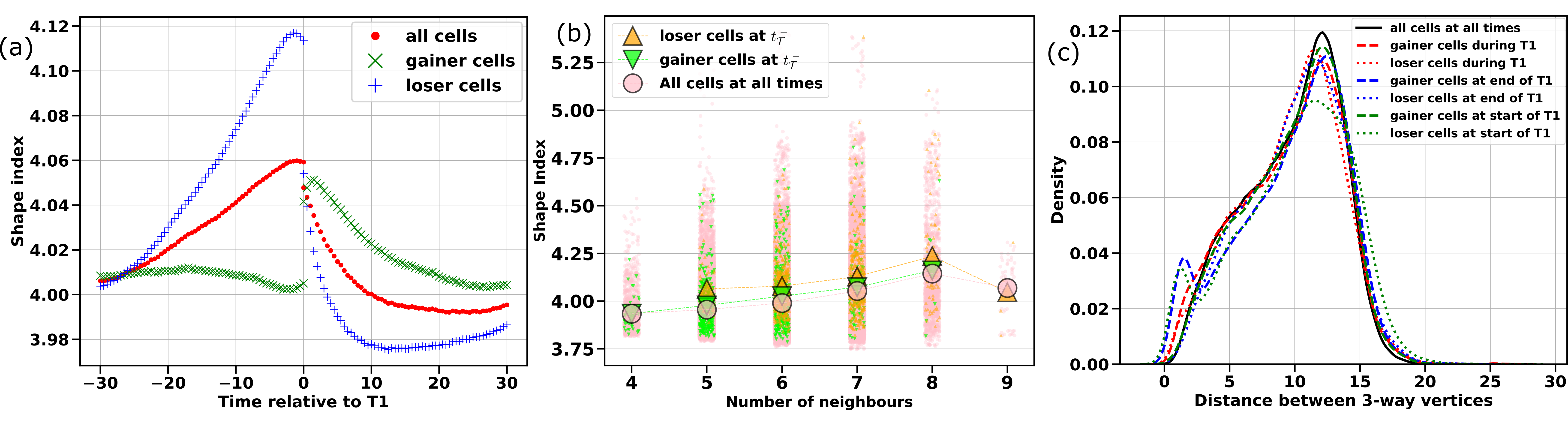}
    \caption{For the MPF model: (a) Evolution of mean shape index of cells undergoing a T1 transition. Negative time and positive time corresponds to relative time before the start of a T1 transition $t_{\mathcal{T}^{-}}$, and relative time after the end of a T1 transition $t_{\mathcal{T}^{+}}$. The mean evolution of gainer and loser cells are also highlighted. (b) Shape index is plotted against the number of neighbours of a cell (blue). The shape index for loser and gainer cells at the start of the T1 transition are also shown. The smaller circles mark individual data, while the larger circles represent the mean value calculated for a particular number of neighbours. (c) Probability density of distance between all pairs of vertices of a cell in various scenarios. The black curve corresponds to distance between all pairs of vertices of a cell, ensembled over all cells at all times. The red curves corresponds to the time when the cell is undergoing a T1 transition($t_{\mathcal{T}}^- < t < t_{\mathcal{T}}^+$), whereas the blue and green curves corresponds to the times $t_{\mathcal{T}}^-$ and $t_{\mathcal{T}}^+$ respectively. Dashed and dotted curves represent to gainer or loser cells, respectively. See Movie 3 in the Supplemental Material for movie of the the evolution of the shape of a cell along with its shape index and number of neighbours.}
    \label{fig:4}
\end{figure*}
The cell shape index $p$ defined as 
\begin{equation}
    p = \frac{\text{Perimeter}}{\sqrt{\text{Area}}},
\end{equation}
is a common cell shape descriptor especially for quantifying cross-over regimes in jamming transitions~\cite{bi2016motility}. For instance, in asthmatic airway epithelium, the jamming transition has been associated to a critical shape index of $p\approx 3.81$, below which the cells are jammed~\cite{Park2015}. 

In our MPF model, the cell area remains invariant, meaning that the shape index is determined solely by the cell perimeter. Since the energy is concentrated at the cell-cell junctions, a cell with a higher perimeter (or shape index) possesses a higher associated energy. Therefore, the shape index also serves as an indicator of the energy associated with the shape of a cell.

In Figure~\hyperref[fig:4]{4a}, we show the time evolution of the shape index of cells that take part in T1 transitions, averaged over all T1 transitions, as function of the time before and after a T1 transition. The evolution after a T1 transition is shown such that  $t_{\mathcal{T}}^+$ is fixed at zero, with positive time representing the period after $t_{\mathcal{T}}^+$. Similarly, the time evolution before a T1 transition is shown such that $t_{\mathcal{T}}^-$ is fixed at zero, with negative time representing the period before $t_{\mathcal{T}}^-$~\cite{jainRobustStatisticalProperties2023}. The average shape index of cells in T1 transitions increases before the start of T1, reaches its peak at the start of T1, and falls sharply immediately after the end of T1, followed by a slower relaxation.

However, when plotting the evolution separately for \emph{loser} and \emph{gainer} cells, different patterns become evident. Before a T1 event, the shape index of loser cells increases significantly upon elongation, reaching its peak just before the start of the T1 transition. Meanwhile, the shape index of gainer cells remains constant on average. As the four cells are on the verge of undergoing a T1 event, the gainer cells move towards each other, compressing the loser cells such that the junction shared by the loser cells shrinks~\cite{jainRobustStatisticalProperties2023}. Gainer cells elongate during the T1 transition and reach their maximum shape index after the end of the T1 transition. Whereas, the shape index of loser cells reduce during the T1 transition. At the end of the T1 transition, loser and gainer cells have similar shape indices. After the T1 transition, the shape index of loser cells decays at a faster rate than that of the gainer cells. Thus, the loser and gainer cells experience distinctive shape deformations due to a T1 transition.

To gain more insights, we plot the distribution of the shape index conditional on the number of neighbours a cell has, for all cells at all times, as shown in Figure~\hyperref[fig:4]{4b}. The figure also shows the shape index of \emph{loser} and \emph{gainer} cells at the start of a T1 transition. The mean values, indicated by larger markers, show that the average shape index tends to increase with the number of neighbours. As cells with fewer neighbours have a smaller shape index, T1 transitions lead to shape relaxation of loser cells. Similarly, T1 transitions cause elongation of gainer cells, leading to shape tension. This is despite the fact that loser cells undergo more drastic deformations than gainer cells, as evident from the evolution of shape index in Figure~\hyperref[fig:4]{4a}. 

Cells with higher activity $v_0$ undergo more T1 transitions and have a higher mean shape index~\cite{jainRobustStatisticalProperties2023}. As the mean shape index increases with activity and as cells undergo more T1s, they develop transient elongations, or protrusions. 

Figure~\hyperref[fig:4]{4c} shows the distribution of distances between pairs of cell vertices. The black curve shows the distribution of distances between all pairs of cell vertices for all cells at all times. The other curves are conditional on certain situations a cell is in. 
\begin{itemize}
    \item The red curve shows the distances when the cell is undergoing a T1 transition, during the period $t_{\mathcal{T}}^- < t < t_{\mathcal{T}}^+ $.

   \item The green curve corresponds to the distances at the start of a T1 transition, $t_{\mathcal{T}}^-$.

   \item The blue curve represents the distances at the end of a T1 transition $t_{\mathcal{T}}^+$.
\end{itemize}
Overall, the curves are similar, but there is a noticeable small peak at short distances for loser cells at the beginning of a T1 transition and for gainer cells at the conclusion. These peaks indicate that the pair of vertices that are about to annihilate or have recently been created. When two vertices of a cell are close to each other, the shape of the cell is likely very anisotropic (elongated). This explains why the shape index of loser cells peaks before the T1 transition, and the shape index of gainer cells peaks afterward. 
\section{Emergent directional migration}\label{sec: successive T1s}
\begin{figure*}
    \centering
    \includegraphics[width=\textwidth]{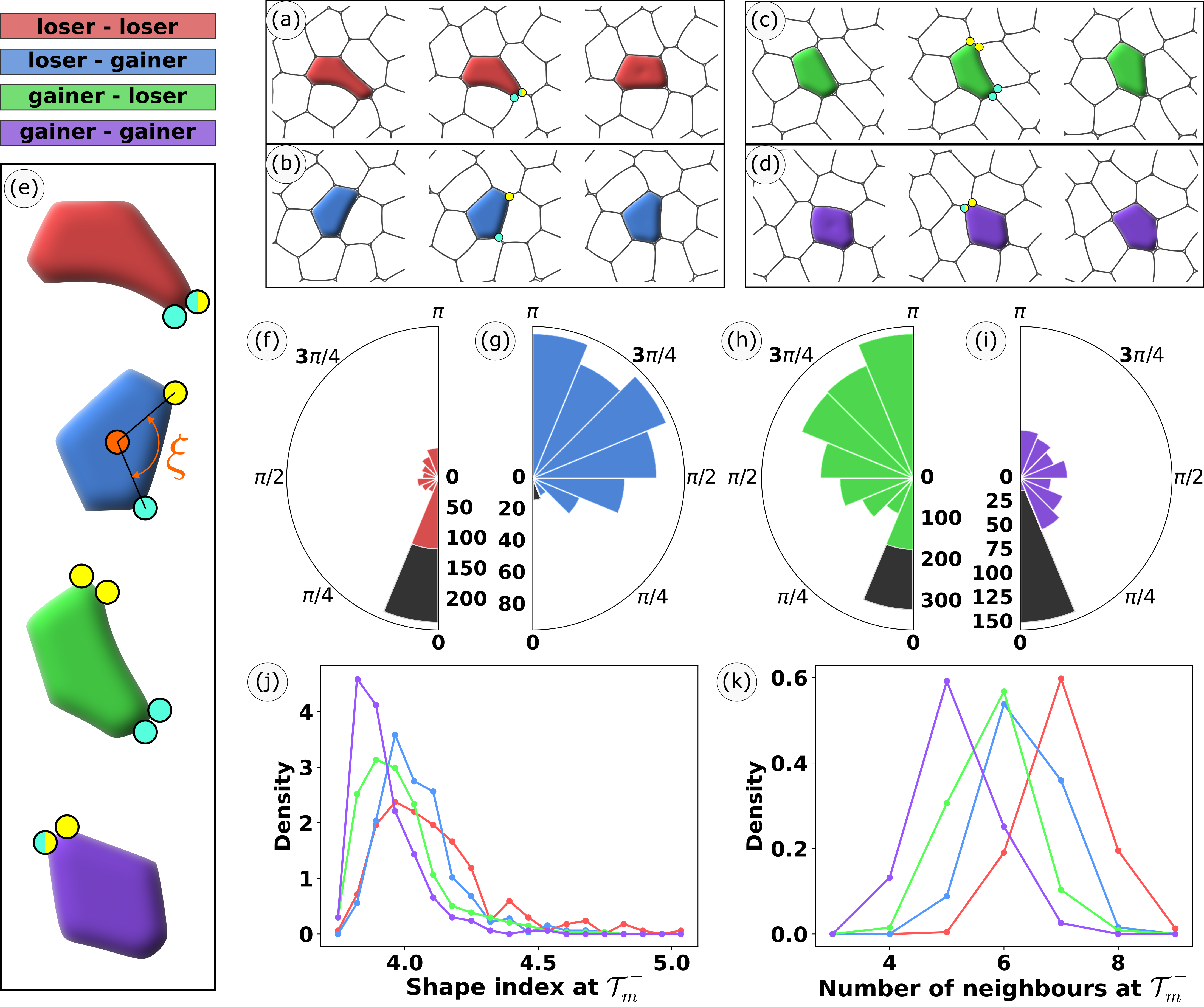}
    \caption{For the MPF model: (a-d) Four different patterns by which highlighted cell $i$ undergoes two \emph{successive} T1 events, i.e. $\mathcal{T}_m$($1^{\text{st}}$ T1) and $\mathcal{T}_n$($2^{\text{nd}}$ T1). The colours correspond to legend at the top-left. $\mathcal{T}_m$ takes places between the left and the middle column, while $\mathcal{T}_n$ takes places between the middle and the right column. The yellow and cyan circles mark the vertices of the highlighted cell that are created in $\mathcal{T}_m$ and annihilated in $\mathcal{T}_n$, respectively. Sometimes, vertices take part in both the T1s, and are marked with both cyan and yellow. (a) \emph{loser-loser}: Red cell loses neighbours in both T1s. (b) \emph{loser-gainer}: Blue cell loses neighbours in $\mathcal{T}_m$ and gains neighbours in $\mathcal{T}_n$. (c) \emph{gainer-loser}: Green cell gains neighbours in $\mathcal{T}_m$ and loses neighbours in $\mathcal{T}_n$. (d) \emph{gainer-gainer}: Purple cell gains neighbours in both T1s. (e) Shows the zoomed highlighted cells from (a-d). (f-i) Fixing the origin to the center of mass of cell $i$, the difference in orientation of each element in the set of vertices created in ($\mathcal{T}_m$) (i.e., ($\mathcal{U}_{i}^+(\mathcal{T}_m))$) with each element in the set of vertices annihilated in ($\mathcal{T}_n$) (i.e., ($\mathcal{U}_{i}^-(\mathcal{T}_n))$) is histogrammed over all pairs of \emph{successive} T1 transitions for all cells. An example of the difference in orientation indicated by $\xi$ on the blue cell, as shown in (e). $\xi$ is the difference in orientation of blue and yellow circles measured relative to the orange circle(center of mass). The black colour in (f-i) represents vertices that are created in $\mathcal{T}_m$ and annihilated in $\mathcal{T}_n$. (j) Shape index and (k) number of neighbours of the cell undergoing \emph{successive} T1s measured at the start of $\mathcal{T}_m$, with the color marking the appropriate scenario. The statistics in (f-k) is collected over three simulations.}
    \label{fig:5}
\end{figure*}
Since cell shape and its tendency to lose or gain neighbours during a T1 transition are interconnected, we now investigate whether the interplay between cell shape and junction remodeling causes coherent flow patterns that span larger length scales than those typical of a T1 transition.

Let $\mathcal{U}_{i}^-(\mathcal{T})$ and $\mathcal{U}_{i}^+(\mathcal{T})$ be the set of all vertices that were annihilated and created, respectively, due to the T1 transition $\mathcal{T}$ that involved cell $i$.

For a pair of \emph{successive} T1s $\mathcal{T}_m$ and $\mathcal{T}_n$ involving cell $i$, i.e., $i \in \mathcal{T}_m$ and $i \in \mathcal{T}_n$, the following relations hold. 
\begin{equation*}
     \begin{array}{rl}
     \mathcal{U}_{i}^+(\mathcal{T}_m)=\{\mathcal{X}_{\mathcal{T}_m, c}^+\}&: \;\; \text{if $i$ is a loser cell in $\mathcal{T}_m$}\\  \mathcal{U}_{i}^+(\mathcal{T}_m)=\{\mathcal{X}_{\mathcal{T}_m, 1}^+, \mathcal{X}_{\mathcal{T}_m, 2}^+\} &: \;\; \text{if $i$ is a gainer cell in $\mathcal{T}_m$} \\
          \mathcal{U}_{i}^-(\mathcal{T}_n)=\{\mathcal{X}_{\mathcal{T}_n, 1}^-, \mathcal{X}_{\mathcal{T}_n, 2}^-\} &: \;\; \text{if $i$ is a loser cell in $\mathcal{T}_n$} \\ \mathcal{U}_{i}^-(\mathcal{T}_n)=\{\mathcal{X}_{\mathcal{T}_n, c}^-\}&: \;\;  \text{if $i$ is a gainer cell in $\mathcal{T}_n$}
     \end{array} 
\end{equation*}
where $c=1$ or $2$ depending on the vertex which includes cell $i$.

There are four different scenarios for a cell involved in \emph{successive} T1 transitions, as shown in the Figure~\hyperref[fig:5]{5a-5d}, depending on whether a cell is a loser or a gainer cell in the two T1 transitions. These four scenarios are denoted as \emph{loser-loser}, \emph{loser-gainer}, \emph{gainer-loser}, \emph{gainer-gainer}. The vertices of the highlighted cell involved in $\mathcal{T}_m$ and  $\mathcal{T}_n$ are marked in yellow and cyan, respectively. In some situations, the same vertex can be involved in both T1 transitions, such that it is created in $\mathcal{T}_m$ and annihilated in $\mathcal{T}_n$, and is appropriately highlighted using both cyan and yellow. This is possible in all four scenarios of \emph{successive} T1 transitions. A cell undergoes a reversible T1 transition in the scenarios \emph{loser-gainer} and \emph{gainer-loser} only if both the vertices involved in the $\mathcal{T}_m$ are involved in $\mathcal{T}_n$. We do not encounter any reversible T1 transitions in the MPF models in the parameter regime we have explored, but we have encountered several T1 transitions where one vertex was involved in \emph{successive} T1 transitions.

For each of the four scenarios of \emph{successive} T1s involving cell $i$, we first determine the locations of all vertices created after $\mathcal{T}_m$ i.e. $\{\mathbf{r}_{\mathcal{X}}(t_{\mathcal{T}_n}^-)\mid \mathcal{X} \in \mathcal{U}_i^+(\mathcal{T}_m)\}$, and the locations of all vertices that will annihilate in $\mathcal{T}_n$, i.e. $\{\mathbf{r}_{\mathcal{X}}(t_{\mathcal{T}_n}^-)\mid \mathcal{X} \in \mathcal{U}_i^-(\mathcal{T}_n)\}$, both calculated at the start time of $\mathcal{T}_n$ i.e $t_{\mathcal{T}_n}^-$, which is the latest time where the vertices in both $\mathcal{U}_i^+(\mathcal{T}_m)$ and $\mathcal{U}_i^-(\mathcal{T}_n)$ coexist. Then, we determine the location of the center of mass of the cell at $t_{\mathcal{T}_n}^-$ i.e. $\mathbf{r}_i(t_{\mathcal{T}_n}^-)$(calculated using Equation~\ref{eq:COM}). We pair every vertex of   $\mathcal{U}_i^+(\mathcal{T}_m)$ with every vertex of $\mathcal{U}_i^-(\mathcal{T}_n)$. By pairing vertices in this way, in the scenarios \emph{loser-loser}, \emph{loser-gainer}, \emph{gainer-loser} and \emph{gainer-gainer}, we get 2, 1, 4 and 2 pairs of vertices for each pair of \emph{successive} T1s, respectively. 

We fix an origin at the center of mass of the cell, denoted by $\mathbf{r}_i(t_{\mathcal{T}_n}^-)$. For each pair of chosen vertices, we find the difference in orientations of both vertices measured relative to the cell center, which we will call \emph{relative orientation}. An example of such a relative orientation is highlighted by the angle $\xi$ on the blue cell in Figure~\hyperref[fig:5]{5e}. This process is repeated for all pairs of \emph{successive} T1s and for all cells, and the corresponding histograms are shown in the Figures~\hyperref[fig:5]{5f-5i} for all four scenarios. The relative orientation between each vertex pair varies from $0$ to $\pi$. If the same vertex belongs to both $\mathcal{U}_i^+(\mathcal{T}_m)$ and $\mathcal{U}_i^-(\mathcal{T}_n)$, then the data is highlighted in black, and the relative orientation between such a pair is $0$. If both the vertices are on the opposite side of the cells then the relative orientation between such a pair is $\pi$. In essence, the distributions in Figures~\hyperref[fig:5]{5f-5i} informs on the spatial distribution of \emph{successive} T1s with respect to the cell center. Figure~\hyperref[fig:5]{5j} and Figure~\hyperref[fig:5]{5k} show the corresponding probability density distributions of the shape index and the number of neighbours of the cell in each of the four scenarios measured at the start of $\mathcal{T}_m$. 

For the scenario where a cell loses neighbours in both T1 transitions, i.e., \emph{loser-loser}, we see that the distribution peaks at a low value of relative orientation. This suggests that the pairs of vertices that took part in \emph{successive} T1s are close to each other. A large fraction of these pairs involve one vertex taking part in both T1s. Very few pairs of vertices lie on the opposite sides of the cells suggests that most pairs of \emph{successive} T1s happen close to each other on the same side of the cell. Hence, elongated cells are more likely to undergo \emph{successive} T1s of \emph{loser-loser} kind, as shown in Figure~\hyperref[fig:5]{5j}. An elongated cell can relax its shape by retracting its mass towards the cell center along the direction of elongation through T1 transitions (see red cell in Figure~\hyperref[fig:5]{5a}).  Cells in this scenario also have a higher than average number of neighbours as shown in Figure~\hyperref[fig:5]{5k}. 

For the scenario \emph{loser-gainer}, both T1s happen mostly on the opposite sides of a cell as the relative orientation between most pairs is greater than $\pi/2$, as shown in Figure~\hyperref[fig:5]{5g}. This is akin to a cell retracting mass from one side and moving in the opposite direction. The loser cells which undergo more drastic shape changes during a T1, as shown in Figure~\hyperref[fig:4]{4a}, are propelled away from the epicenter~\cite{jainRobustStatisticalProperties2023}, and such a cell gains a neighbour on the opposite side. 

In the scenario \emph{gainer-loser}, we see two peaks near $0$ and $\pi$, see Figure~\hyperref[fig:5]{5g}. The differences in the distributions for \emph{loser-gainer} and \emph{gainer-loser} suggest that the spatial distribution of \emph{successive} T1s on a cell depends on the order of topological change, even if the net topological change in both scenarios is the same. This might be because gainer and loser cells undergo distinct deformations due to T1 transitions, as shown in Figure~\hyperref[fig:4]{4a}. In both \emph{loser-gainer} and \emph{gainer-loser} situations, at the start of the first T1, the average number of neighbours is around 6, suggesting that these two scenarios involve fluctuations about the mean number of neighbours which is approximately 6. 

For the scenario \emph{gainer-gainer}, most pairs of vertices of \emph{successive} T1s lie on the same side of the cell, and most of these also involve the same vertex in both T1s, as shown in Figure~\hyperref[fig:5]{5i}. If we were to exclude the situations where the same vertex is involved, the remaining pairs are likely to occur at any relative orientation except for very low relative orientation, as indicated by the histogram. Cells that gain two neighbours in \emph{successive} T1s usually have a lower shape index at the start of the first T1 than the other three scenarios, as shown in Figure~\hyperref[fig:5]{5j}, as well as a lower number of neighbours, as shown in Figure~\hyperref[fig:5]{5k}. 

If the distributions of relative orientation of the vertices involved in \emph{successive} T1 transitions were uniform, then T1 transitions would have no preferred orientation for the quadrupolar flow field that arises due to them~\cite{jainCellIntercalationFlow2024a}. However, the non-uniform distributions in Figures~\hyperref[fig:5]{5f-5i} suggest an effective directed migration of the cells, depending on the four scenarios, which could lead to emergent flow patterns at larger scales as a result of chaining of T1 transitions. 

The non-uniform spatial distribution of paired vertices of \emph{successive} T1 transitions on the cell might be a reason behind the weak polar order within the MPF model, which is typically more than $0.4$~\cite{jainCellIntercalationFlow2024a}. The large number of paired vertices of \emph{successive} T1 transitions occuring at low relative orientation suggest that the \emph{successive} T1 transitions occur close to each other, informing the profile of the radial distribution functions in Figure~\ref{fig:3} for the MPF model, where a peak was observed for very low value of radial distance ($r<5$). 

In the MPF model, the orientation of cell activity tends to align with the direction of its elongation, and this rate is controlled by a parameter $\alpha$ (See appendix~\ref{App:MPF} for details). However, this alignment is not the reason for the specific spatial distribution of \emph{successive} T1 transitions, as similar distributions are observed even when the activity does not align with cell elongation, i.e., when $\alpha=0$ (See Figure~2 in the Supplemental Material for corresponding plots). Also, these distributions remain similar even if activity and deformability are varied suggesting a universal emergent behaviour due to the shape response of cells. 

Additionaly, we note that every cell that gains neighbours in two \emph{successive} T1s where the vertex created in the first T1 is annihilated in the second T1, also has a neighbouring cell that loses neighbours in both of these T1s. Consider a T1 $\mathcal{T}_m=\{A, B, C, D\}$ where cell $A$ gains a neighbour $B$. This process creates two vertices: $\{A, B, C\}$ and $\{A, B, D\}$. If one of these vertices is involved in another T1 $\mathcal{T}_n$ by annihilating such that cell $A$ gains another neighbour, then depending on which vertex took part in the T1, either $C$ or $D$ loses contact with $B$. So, one of $C$ or $D$ loses neighbours in both $\mathcal{T}_m$ and $\mathcal{T}_n$.

\section{Conclusions:} \label{Sec:conclusions}
In summary, we have shown that while cells undergo individual T1 transitions as a Poisson process in time, with a rate determined by cell activity, their spatial distribution is highly correlated. Through detailed statistical analysis of the cell shape index and the number of cell neighbours, we have found that the cell shape deformation in response to T1 transitions is contingent on whether a cell loses or gains neighbours. This finding highlights a dual role of T1 transitions: cells losing neighbours experience shape relaxation with a decrease in the shape index, whereas cells gaining neighbours become elongated, leading to shape tension.

We identify four distinct occurrence patterns of \emph{successive} T1 transitions. These non-uniform distributions across cell junctions lead to cell shape protrusions or retractions, allowing cells to move directionally relative to others, potentially forming flow structures across multiple cells. However, to better understand the emergent fluidisation patterns, longer chains of successive T1 events need to be considered. The spatial correlations of T1 transitions may also be probed using other models, such as the vertex model~\cite{altVertexModelsCell2017} especially those that allow for complex cell shape deformations as in the Reference~\cite{kimEmbryonicTissuesActive2021a}. These results are also accessible for experimental verification.

\acknowledgements This project is partly funded by the European Union’s Horizon $2020$ research and innovation programme under the Marie Skłodowska-Curie grant agreement No 945371, and by the University of Oslo, through UiO:Life Science through the convergence environment 
ITOM. LA acknowledges support in part by grant NSF PHY-2309135 to the Kavli Institute for Theoretical Physics (KITP) and the Gordon and Betty Moore Foundation Grant No. 2919.02.

\appendix 

\section{Multi-phase field model}\label{App:MPF}
The multi-phase field model presented here was previously used in Ref.~\cite{jainCellIntercalationFlow2024a}, and is based on earlier models in Refs~\cite{wenzelTopologicalGeometricalQuantities2019, jainRobustStatisticalProperties2023}. Consider a set of cells represented by phasefields $\{\phi_i\}$ defined in $\Omega \subset \mathbb{R}^2$, where $i=1, 2, \dots, N$, and $N$ is the total number of cells. The domain $\Omega$ is a square of side length $L$. The interior and exterior of the cell $i$ is identified as the region in $\Omega$ with bulk phase values $\phi_i\approx 1$ and  $\phi_i\approx -1$, respectively. The interior is phase separated from the exterior through a diffuse interface of width $\mathcal{O}(\epsilon)$. The phase fields $\phi_i$ evolve according to 
\begin{equation}\label{eq:system MPF}
    \partial_t \phi_i + \mathbf{v}_i \cdot \nabla \phi_i = \Delta \frac{\delta  \mathcal{F}}{\delta \phi_i}, 
\end{equation}
while preserving the total cell area. The free energy functional $\mathcal{F}[\{\phi_i\}] = \mathcal{F}_{CH}[\{\phi_i\}] + \mathcal{F}_{INT}[\{\phi_i\}]$ is a sum of energy of the surface and the interaction energy between the cells. The energy of the surface is the Cahn Hilliard energy given as 
\begin{equation}\label{eq: Cahn Hilliard Energy}
    \mathcal{F}_{CH} [\{\phi_i\}]= \frac{1}{Ca}\sum_{i=1}^N \int_{\Omega} g(\phi_i)\left(\frac{\epsilon}{2}||\nabla \phi_i||^2 + \frac{1}{4\epsilon}(\phi_i^2-1)^2\right) dx,
\end{equation}
where $Ca$ is the capillarity number that controls the deformability of the cells, $\epsilon$ controls the thickness of cell interface, and $g(\phi)=\frac{2}{3(\phi_i+1)(\phi_i-1)}$ is the diffusion restriction function to prevent bulk diffusion~\cite{salvalaglioDoublyDegenerateDiffuse2020}. The first term penalises gradients in the phase field which models cortical tension, while the second term is the demixing part which ensures phase separation between the exterior and interior of the cell. The interaction energy is given as 
\begin{dmath}
    \mathcal{F}_{INT}[\{\phi_i\}] = \frac{1}{In}\sum_{i=1}^N \int_{\Omega}\sum_{j\neq i} \left(\frac{a_r}{2}(\phi_i+1)^2(\phi_j+1)^2-\frac{a_a}{2}(\phi_i^2-1)^2(\phi_j^2-1)^2\right)dx 
\end{dmath}

where, $a_r$ and $a_a$ are used to tune the repulsive and attraction component, respectively. The repulsive part penalises overlap of cell interiors, while the attractive part promotes overlap of cell interfaces. 

The cell activity field is given as 
\begin{equation}
    \mathbf{v}_i(\mathbf{x}, t) = v_0 \mathbf{e}_i(t)\hat\phi_i(\mathbf{x},t)
\end{equation}
where $v_0$ is the magnitude of activity, $\hat{\phi}=0.5\phi+0.5$, and $\mathbf{e}_i= [\cos \theta_i (t), \sin  \theta_i(t) ]$ is the activity orientation vector. The orientation angle $\theta$ evolves according to the stochastic differential equation 
\begin{equation}
    d\theta_i = \sqrt{2 D_r }dW_i(t) + \alpha(\beta_i(t) - \theta_i(t))dt,
\end{equation}
where $D_r$ is the coefficient of rotational diffusion and $W_i$ is a Wiener process. The angle $\beta_i(t)$ is the angle of shape polarity defined to be the elongation axis of the cell, and is calculated from the eigenvector corresponding to the largest eigenvalue of the shape deformation tensor, $\mathbf{\eta}_i^+$ such that 
\begin{equation}
     \beta_i(t) = \left\{ 
 \begin{array}{rl}
 \arg(\boldsymbol{\eta}_i^{+}(t)) &: \;\; \mathbf{e}_i(t)\cdot \boldsymbol{\eta}_i^{+}(t) > 0 \\ \arg(-\boldsymbol{\eta}_i^{+}(t)) &: \;\; \mathbf{e}_i(t)\cdot \boldsymbol{\eta}_i^{+}(t) < 0 .
 \end{array}\right.
\end{equation}
The system of partial differential equations in Equation~\ref{eq:system MPF} is considered on a square periodic domain. The system is spatially discretised using finite elements with a semi-implicit discretisation in time, and is solved using adaptive finite element methods using the C++ FEM tool AMDIS~\cite{Vey2007, Witkowski2015} with the parallisation approach introduced in Reference~\cite{praetoriusCollectiveCellBehaviour2018}. For each simulation, we initialize the system with $N$ square cells arranged in a regular grid with randomly oriented activity, allowing for an initial transient period for cells to mix. We consider the data only after the cell shapes appear sufficiently irregular.

The parameters used are as per Table~\ref{tab:parameters} unless specified otherwise.
    \begin{table}[htb!]
    \centering
    \begin{tabular}{|c|c|c|c|c|c|c|c|c|c|c|c|}
        \hline
        $N$ & $L$ & $T$ & $\epsilon$ & $v_0$ & $Ca$ & $In$ & $a_a$ & $a_r$ & $D_r$ & $\alpha$ & $dt$ \\
        \hline
        $100$ & 100 & 150 & 0.15 & 0.5 & 0.2 & 0.1 & 1 & 1 & 0.01 & 0.1 & 0.005 \\
        \hline
    \end{tabular}
    \caption{Default values of the model parameters. $dt$ denotes the timestep size.}
    \label{tab:parameters}
    \end{table}

The center of mass of a cell $i$ in MPF model is calculated as
\begin{equation}\label{eq:COM}
    \mathbf{r}_i(t) = \frac{\int_{\Omega}\mathbf{x}\hat{\phi_i}dx}{\int_{\Omega}\hat{\phi_i}dx}
\end{equation}
where $\hat{\phi_i} = 0.5\phi_i+0.5$ is the rescaled phasefield. 

\section{Active elastic solid model}\label{App:ASM}

The active elastic solid model~\cite{ferrante2013elasticity} is a discrete particle model consisting of a collection of $N$ particles at positions $\mathbf r_i$ in a flat tissue monolayer. The elasticity of the tissue monolayer is modelled by connecting neighbouring cells with spring-like forces $\mathbf f_i = -\nabla_{\mathbf r_i} V$ determined by a generic Gaussian potential for the elastic interactions, 
\begin{eqnarray}
V &=& \sum\limits_{i}\sum\limits_{j= n.of.i} \exp\left\{-\frac{\left(|\mathbf r_{ij}|-l_{ij}\right)^2}{a_0^2}\right\}
\end{eqnarray}
where $\mathbf r_{ij} = \mathbf r_{i}-\mathbf r_{j}$ is the separation vector between two cells and $l_{ij}$ is the equilibrium distance between cells. The sum of $j$ is taken over the neighbours on an initially square lattice. The directional self-propulsion follows the cell polarities $\mathbf p_i$ which reorients due to a torque induced by elastic interactions. The overdamped dynamics on this system is then described by   
\begin{eqnarray}
    \dot{\mathbf r}_i &=& v_0 \mathbf p_i + \frac{1}{\zeta} \left(\mathbf f_i +\xi_i\right)\\ 
    \dot{\mathbf p}_i &=&  \mathbf T_i \times \mathbf p_i + \eta_i\ .
\end{eqnarray}
where the polarity torque is 
\[\mathbf T_i = \frac{1}{\zeta}(\mathbf p_i \times \dot{\mathbf r}_i) \ .\]
The translational noise has zero mean and  
\begin{equation}
    \langle\xi_i(t)\cdot\xi_j(t')\rangle = D_r \delta_{ij}\delta(t-t') 
\end{equation}
Similarly, the rotational noise also has zero mean and 
\begin{equation}
    \langle\eta_i(t)\cdot\eta_j(t')\rangle = D_p \delta_{ij}\delta(t-t') \ .
\end{equation}
We renormalize the magnitude of polarity to unity at every timestep because the noise is not norm conserving.
AES simulations are for 576 cells on a periodic domain of length $L$ with time advancement using the Euler method and with the following parameters unless otherwise given on a figure.

    \begin{table}[htb!]
    \centering
    \begin{tabular}{|c|c|c|c|c|c|c|c|}
        \hline
        $L$ & $T$ & $l_{ij}$ & $1 / a_0^2$ & $\zeta$ & $D_r$ & $D_p$ & $dt$ \\
        \hline
        40 & 5000 & 2 & 0.08 & 1 & 0.025 & 0.15 & 0.1 \\
        \hline
    \end{tabular}
    \caption{Default values of the model parameters for AES. $dt$ denotes the timestep size.}
    \label{tab:parametersAES}
    \end{table}

Local cell rearrangements are imposed by a geometric rule for nearest neighbours. Namely, cells $i$ and $j$ are considered neighbours if and only if they are closest in
distance to their midpoint than any other third cell.
Explicitly, we connect all cells $i$ and $j$ and remove the connection if there exists any third cell $k$ which satisfies $r_{ij}^2 > r_{ik}^2 + r_{jk}^2$, with $r_{ij}$ being the distance between cells $i$ and $j$.
This is the same neighbour criterion as in Reference~\cite{ho2024role}. 
The neighbour criterion is slightly different than using a Voronoi tesselation.
Whilst we do gain information about the duration of T1 transitions, since they no longer occur instantaneously, we do not gain any information about the shape of cells. 
Because of this, we use the multiphase field model to answer these questions.

In the simulations used in this paper, since we are concerned with the behavior of T1s specifically, we update the neighbour list at every time-step.
We then define a T1 as occuring in the following way: when cells $i$ and $j$ cease to be neighbours this is the start of a potential T1; a T1 occurs when cells $m$ and $n$ become neighbours, where both $m$ and $n$ are neighbours of both $i$ and $j$; it is cancelled if $i$ and $j$ become neighbours again.
According to these rules rosettes are counted as one T1 transition.
The neighbour criterion induces changes in the neighbour lists with time.

\bibliography{t1chains}
\end{document}


\preprint{APS/123-QED}


\title{Supplementary Material}

\author{Harish P. Jain}
\affiliation{
Njord Centre\\ Department of Physics\\ University of Oslo}
 
\author{Richard D.J.G. Ho}
\affiliation{
Njord Centre\\ Department of Physics\\ University of Oslo}

\author{Luiza Angheluta}
\email{luiza.angheluta@fys.uio.no}
\affiliation{
Njord Centre\\ Department of Physics\\ University of Oslo}

\maketitle
\section{Additional figures}
\begin{figure}[h]
    \centering
    \includegraphics[width=0.7\linewidth]{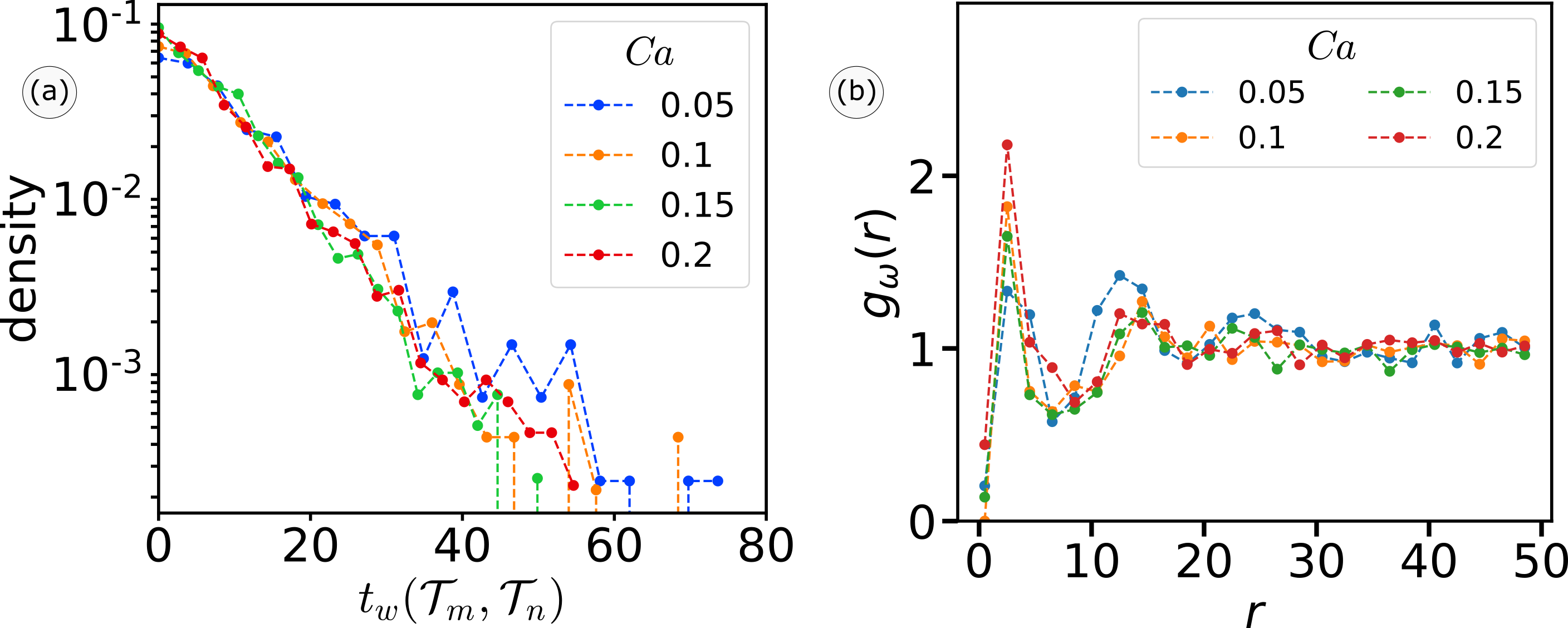}
    \caption{\textbf{Spatiotemporal properties for varying deformability $Ca$ in the MPF model}: (a) Waiting time $t_w$ probability density distribution between \emph{successive} T1s $\mathcal{T}_m$ and $\mathcal{T}_n$ undergone by a cell in the MPF model for varying deformability $Ca$. (b) Radial distribution function $g_{\omega}(r)$ between T1 transitions for MPF model for $\omega=25$ and varying deformability $Ca$.}
    \label{fig:supplementary1}
\end{figure}

\begin{figure}[h]
    \centering
    \includegraphics[width=0.9\linewidth]{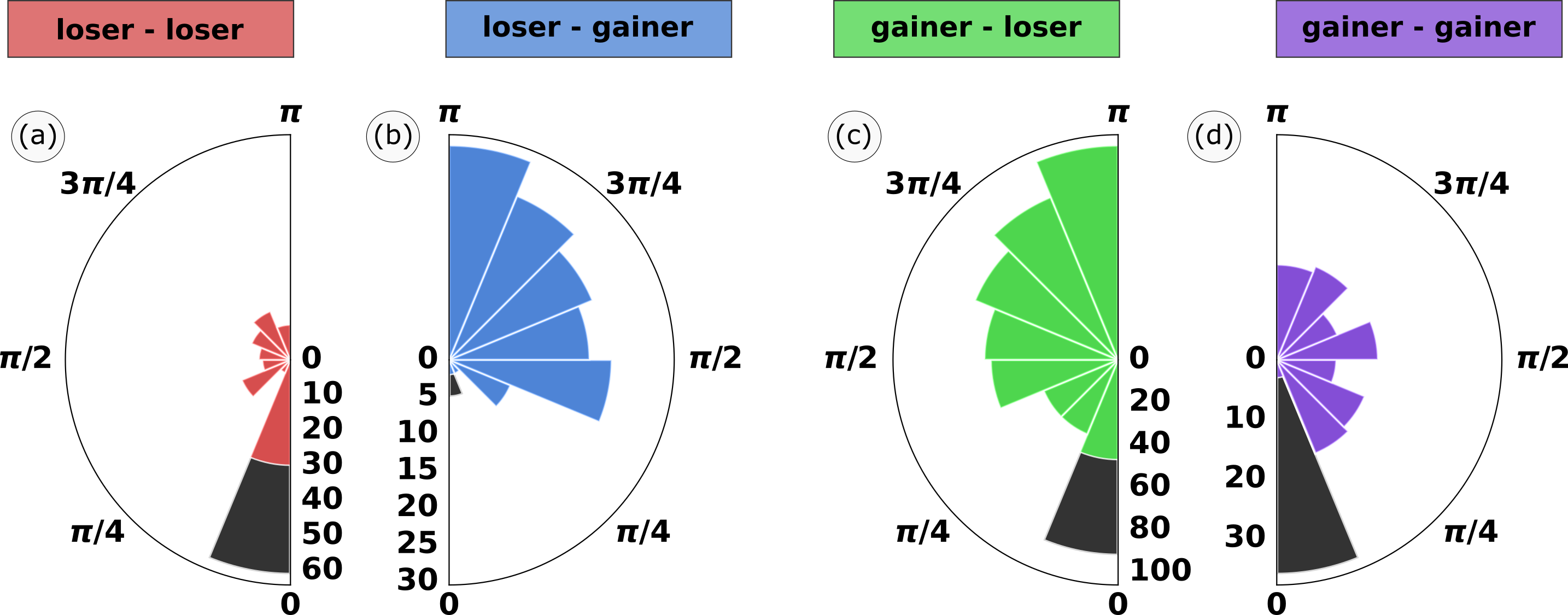}
    \caption{\textbf{Successive T1s in absence of shape alignment $\alpha=0$}: (a-d) For four different patterns of \emph{successive} T1 events, i.e., $\mathcal{T}_m$($1^{\text{st}}$ T1) and $\mathcal{T}_n$($2^{\text{nd}}$ T1) undertaken by a cell: Fixing the origin to the center of mass of cell $i$, the difference in orientation of each element in the set of vertices created in ($\mathcal{T}_m$) (i.e., ($\mathcal{U}_{i}^+(\mathcal{T}_m))$) with each element in the set of vertices annihilated in ($\mathcal{T}_n$) (i.e., ($\mathcal{U}_{i}^-(\mathcal{T}_n))$) is histogrammed over all pairs of \emph{successive} T1 transitions for all cells. The black colour in represents vertices that are created in $\mathcal{T}_m$ and annihilated in $\mathcal{T}_n$. $\alpha=0$ and other parameters are as per Table I of the main text.}
    \label{fig:supplementary2}
\end{figure}